# Interface Gráfica para Simulação de Amplificadores Óticos: Ferramenta Pedagógica

T. Almeida, R.A.S Ferreira and P. S. André, *Senior Member, IEEE*

*Abstract*— This study presents the development of two graphical interfaces using Matlab's graphical user interface tool, GUIDE, to model the steady-state and dynamic responses of Erbium (Er3+)-doped fiber amplifiers. The primary focus is on their application for educational purposes, employing two distinct models. The steady-state model aims to evaluate the influence of various intrinsic parameters of $Er^{3+}$-doped fibers, such as length, overlap factor between the dopant profile and the optical signal, dopant density, and excited state average lifetime, along with the impact of optical signals on the amplifiers' performance. On the other hand, the dynamic model interface facilitates the examination of the channels add/drop impact in an optical network. This dual-interface approach provides a comprehensive understanding of the Erbium-doped fiber amplifiers, making them valuable tools for educational exploration and analysis.

*Resumo*— Este trabalho apresenta o desenvolvimento de duas interfaces gráficas utilizando a ferramenta de interface gráfica do Matlab, GUIDE, para modelizar as respostas em estado estacionário e dinâmicas de amplificadores de fibra dopada com Érbio (Er3+). O foco principal está na aplicação destas interfaces para fins educacionais, empregando dois modelos distintos. O modelo em estado estacionário visa avaliar a influência de vários parâmetros intrínsecos das fibras dopadas com $Er^{3+}$, tais como comprimento, fator de sobreposição entre o perfil do dopante e o sinal óptico, densidade do dopante e tempo de vida médio do estado excitado, juntamente com o impacto dos sinais ópticos no desempenho dos amplificadores. Por outro lado, a interface do modelo dinâmico facilita a análise do impacto de adição/remoção de canais numa rede óptica. Esta abordagem de dupla interface proporciona uma compreensão abrangente dos amplificadores de fibra dopada com Érbio, tornando-os ferramentas valiosas para exploração e análise educacional.

## I. INTRODUÇÃO

Os amplificadores de fibra ótica dopada com iões Érbio ($Er^{3+}$), designados por AFDE's, são utilizados nos sistemas de comunicações óticas desde o início dos anos 90, do século XX [1]. Atualmente, a sua função como repetidores de sinal em sistemas de longa distância foi alargada para utilização nas redes da nova geração, onde a infraestrutura de fibra ótica está presente até a casa dos subscritores [2].

Nas instituições do ensino superior, os AFDE's, bem como todas as soluções de amplificação ótica (amplificadores de semicondutor e amplificadores de Raman) [3], são abordados em cursos de diversas áreas científicas como, por exemplo, Física, Engenharia Eletrotécnica ou Telecomunicações. Em virtude da multidisciplinaridade do tópico, e da coexistência de alunos com diferentes níveis e enquadramento de formação base, os planos de estudo desses cursos são restritos no ensino detalhado do desempenho destes amplificadores e dos fenómenos físicos associados [4].

Assim, a simulação computacional da resposta de amplificadores óticos é um tópico de relevo, pois permite modelizar as equações que governam o comportamento dos AFDE's e testar um elevado número de cenários de aplicação, de forma acessível e rápida, flexibilizando os métodos de ensino e aprendizagem.

Neste contexto, este trabalho apresenta o desenvolvimento de um conjunto de interfaces gráficas, programadas em Matlab, que podem ser utilizadas em ambiente de sala de aula, promovendo um ensino interativo e eficaz. Mais ainda, permite, também, a sua utilização pelos alunos fora das horas de contacto com o docente [5]. Estas interfaces têm o formato de código aberto e usam funcionalidades do Matlab, que é numa linguagem familiar aos docentes e discentes, e facilita alterações futura, possibilitando uma personalização adequada às necessidades específicas de utilização.

Neste artigo, após uma introdução ao tópico em discussão, na segunda secção, são apresentados os conceitos teóricos relevantes para descrever o funcionamento dos AFDEs. Na terceira secção, descrevem-se os algoritmos utilizados na modelação de AFDE's, sendo, seguidamente, apresentadas as interfaces gráficas desenvolvidas e alguns exemplos de utilização. Por fim, na última secção, elencam-se as conclusões do trabalho.

## II. CONCEITOS TEORICOS

Um amplificador de fibra ótica dopada com iões Érbio ($Er^{3+}$) caracteriza-se por utilizar a fibra ótica como meio ativo para a realização da função de amplificação. A fibra é dopada com iões lantanídeos, neste caso iões $Er^{3+}$, que através do processo de emissão estimulada, transferem a energia do sinal de um laser de bombeamento para uma portadora modulada que transporta a informação a transmitir [6].

Os iões de $Er^{3+}$ são uma boa escolha para a utilização em sistemas de comunicações óticas, pois quando excitados com um sinal ótico com um comprimento de onda de 980 nm ou 1480 nm, emitem na região espectral, usualmente designada como banda espetral C (1530–1565 nm). A banda C é a região

T. Almeida, Instituto de Telecomunicações e Departamento Física, Universidade de Aveiro, telmopelicano@av.it.pt
R. A. S Ferreira, Departamento de Física e CICECO, Universidade de Aveiro,
rferreira@ua.pt
P. S. André, Instituto de Telecomunicações e Departamento de Engenharia Electrótecnica e de Computadores, Instituto Superior Técnico, Universidade de Lisboa, paulo.andre@lx.it.pt

espectral mais utilizada em comunicações óticas por compreender o valor mínimo absoluto de atenuação nas fibras óticas de sílica (em torno de 1550 nm) [6].

A emissão estimulada é o fenómeno responsável pela amplificação do sinal em AFDE's. Para descrever o conceito de emissão estimulada, é necessário, também, referir os processos de absorção estimulada e emissão espontânea. Consideremos um meio onde os átomos são caracterizados por dois estados de energia: o estado fundamental, com energia $E_1$, e o estado excitado, com energia $E_2$ ($E_1 < E_2$). Na presença de radiação com frequência ν, ocorre a absorção do fotão e a transição dos portadores do estado fundamental, $E_1$, para o seu estado excitado, $E_2$, sendo:

$$\upsilon = \frac{E_1 - E_2}{h} \quad (1)$$

onde $h$ é a constante de Planck. Este processo é denominado por absorção espontânea.

Uma vez no estado excitado, um portador pode decair para o estado fundamental e emitir um fotão de frequência ν. Este processo é designado por emissão espontânea, pois ocorre sem interferência externa e é responsável pelo ruído ótico gerado nos AFDE´s (ruído de emissão espontânea amplificada) [6]. No caso de um fotão, induzir o processo de emissão de um outro fotão, que terá a mesma frequência do fotão inicial, o fenómeno é designado por emissão estimulada. Nesta situação, o fotão emitido tem as mesmas características (fase, polarização e frequência) do fotão que induziu o processo.

A emissão estimulada, cujo esquema ilustrativo pode ser visto na figura 1, é o fenómeno chave, que permite a amplificação de um sinal ótico. No entanto, para este processo de emissão ser dominante é necessária a existência de inversão de população, ou seja, a densidade de portadores no estado excitado deve ser superior à densidade de portadores no estado fundamental.

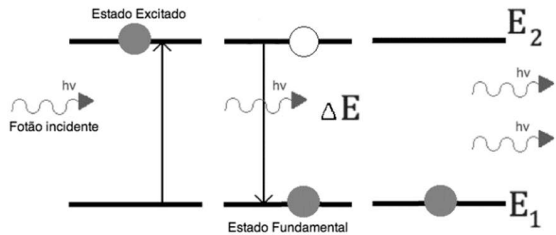

Figura 1. Esquema representativo do fenómeno de emissão estimulada.

Um diagrama genérico dos níveis de energia do ião $Er^{3+}$ na Sílica pode ser visualizado na figura 2.

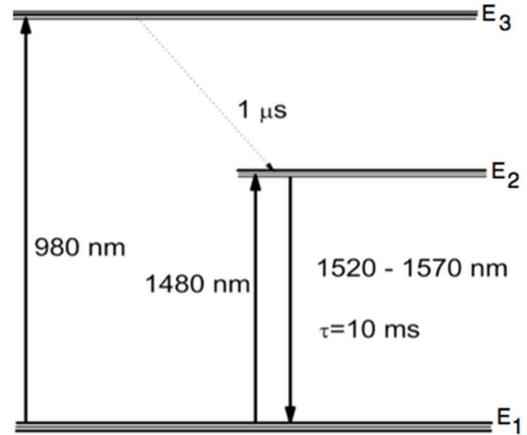

Figura 2. Diagrama genérico de três níveis do ião $Er^{3+}$ [6]. As setas verticais representam (↑) absorção e (↓) emissão. A seta a tracejado representa o decaimento não radiativo entre os níveis $E_3$ ($^4I_{11/2}$) e $E_2$ ($^4I_{13/2}$).

Num AFDE, a amplificação ocorre em vários estágios distintos. Inicialmente, os iões $Er^{3+}$ no seu estado fundamental ($^4I_{15/2}$, designado por $E_1$) são excitados através do bombeamento com um sinal cujo comprimento de onda é 980 nm ou 1480 nm, criando-se uma situação de inversão de população. Caso o sinal de bombeamento tenha com um comprimento de onda de 980 nm, os iões são excitados para um nível de energia superior, ($^4I_{11/2}$, designado por $E_3$) de onde decaem rapidamente (~1 μs), para um estado de menor energia ($^4I_{13/2}$, designado por $E_2$). No caso em que o bombeamento é feito com um sinal cujo comprimento de onda é 1480 nm, os portadores transitam diretamente para o estado $E_2$. O tempo de vida ($\tau$) do estado $E_2$ é de aproximadamente 10 ms (valor típico para iões $Er^{3+}$ em $SiO_2$), que é suficientemente longo para permitir a ocorrência de inversão de população do nível $E_1$ para $E_2$ e, assim, estabelecer as condições para a ocorrência de amplificação ótica [7].

As equações de taxa que descrevem a dinâmica das populações dos AFDEs, utilizadas nas interfaces gráficas desenvolvidas, são baseadas nos modelos propostos por Saleh [8] e Sun [9]. Assumindo que o AFDE pode ser representado por um sistema com dois níveis de energia e com um determinado número de canais a amplificar (sinais que se propagam na fibra com comprimento de onda, $\lambda_k$, e potência ótica, $P_k$,). O meio ativo do AFDE é uma fibra ótica de Sílica dopada com $Er^{3+}$, sendo caracterizada por um comprimento, $L$, uma densidade de dopantes, $\rho$, e uma secção de dopagem, $A$. Segundo o modelo proposto por Saleh, a equação de taxas que descreve a fração de portadores, $N_2$, que ocupa o estado excitado $E_2$ é dada por [8]:

$$\frac{\partial N_2(z,t)}{\partial t} = \frac{N_2(z,t)}{\tau} - \frac{1}{\rho A} \sum_{j=1}^{N} U_j \frac{\partial P_j(z,t)}{\partial z} \quad (2)$$

sendo $N_1(z,t) + N_2(z,t) = 1$. A evolução da potência ótica ao longo da fibra, em função do tempo e para cada canal, $k$, é dada por:

$$\frac{\partial P_k(z,t)}{\partial z} = -\rho U_k \Gamma_k \left[ \left( \sigma_k^e + \sigma_k^a \right) N_2(z,t) - \sigma_k^a \right] \quad (3)$$

sendo $\Gamma$ um fator que contabiliza a sobreposição da distribuição radial dos dopantes com a distribuição modal dos sinais óticos, tomando valores entre 0 e 1. Os termos $\sigma_k^e$ e $\sigma_k^a$ são, respetivamente, os valores das secções cruzadas de emissão e de absorção. Os versores $u_k$ e $-u_k$ indicam o sentido de propagação dos sinais na fibra (1 para sentido co propagante e −1 para o sentido contra propagante).

### III. MODELIZAÇÃO DE AFDES

A maioria dos modelos computacionais disponíveis comercialmente para descrever o funcionamento de AFDE's têm como base as equações de *Saleh* ou de *Sun* (descritas adiante) com aproximações e modificações adaptadas ao problema em questão. Geralmente, os modelos utilizados para descrever os AFDE's podem ser divididos em duas categorias: modelos *i)* estacionários e *ii)* dinâmicos. Nos modelos estacionários considera-se que os sinais não sofrem perturbações ao longo de tempo, ou seja, é eliminada a dependência temporal das equações de taxa. Como tal, estes modelos são os mais indicados para estudar a influência dos parâmetros físicos do amplificador. Sendo, usualmente realizadas aproximações analíticas às equações de taxa, com a aplicação de métodos como o da análise da potência média (APA - *Average Power Analysis*) [10]. Nestas aproximações, exclui-se a influência de parâmetros como a atenuação da fibra ótica dopada e a emissão espontânea amplificada, permitindo desta forma reduzir o tempo de computação.

Nos modelos dinâmicos são consideradas as perturbações temporais do sistema, requerendo para a sua implementação a resolução de equações diferenciais que, geralmente, requerem tempos de computação mais elevados do que aqueles necessários no caso dos modelos estacionários. Os modelos dinâmicos são os mais adequados para o estudo dos AFDE's em redes óticas, especialmente de redes que suportam multiplexagem espectral (WDM-*wave division multiplexing*), onde os sinais óticos são roteados, removidos e adicionados, resultando em perturbações significativas na resposta e desempenho dos AFDE's [11].

Independentemente do tipo de modelo a implementar, é importante chegar a um compromisso entre a complexidade do algoritmo e a versatilidade quando se trata de simular diversos cenários. O objetivo deste trabalho é o de apresentar plataformas de simulação que possam tornar acessível ao utilizador o estudo destes cenários de redes óticas. Para cumprir esse objetivo, foram desenvolvidas duas interfaces para a simulação dos AFDE's, que utilizam um modelo estacionário e um modelo dinâmico de um AFDE. A interface do modelo estacionário permite descrever a influência dos vários parâmetros físicos da fibra dopada e dos sinais óticos envolvidos no desempenho do amplificador e a interface dinâmica permite ilustrar o impacto da adição e remoção de canais no desempenho de um amplificador.

Para implementar a interface do modelo estacionário foi utilizado o método melhorado de análise de potências médias de Hogkinson [10] que tem em conta o fenómeno da emissão espontânea amplificada, permitindo tempos de computação reduzidos (~ms). O modelo tem como base as equações e o formalismo do modelo de Saleh [8] mas elimina a dependência temporal das equações ao assumir que os sinais não sofrem perturbações ao longo do tempo. Este modelo elimina, também, a dependência espacial das equações ao considerar que os dopantes estão uniformemente distribuídos ao longo da fibra. O AFDE é representado como um sistema de dois níveis de energia, $E_1$ e $E_2$, com densidades de portadores, respetivamente, $N_1$ e $N_2$. A evolução da potência ótica dos sinais ao longo da fibra é descrita por [10]:

$$\frac{\partial P_k(z,t)}{\partial z} = -U_k \left[ \left\{ \left(\sigma_k^e + \sigma_k^a\right) \cdot N_2(z) - \sigma_k^a \cdot (N_2 + N_1) \right\} \Gamma_k - \alpha_k \right] P_k^z(z) \quad (4)$$

onde $\alpha_k$ é o termo associado à atenuação da fibra para cada canal. A evolução ao longo da fibra do sinal resultante da emissão espontânea amplificada é dada por (5):

$$\frac{\partial P_i^\pm(z)}{\partial z} = \left[ \left( \Gamma_i \left\{ \left(\sigma_i^e + \sigma_i^a\right) N_2(z) - \sigma_i^a \cdot (N_2 + N_1) \right\} - \alpha_i \right) P_i^\pm(z) + 2\sigma_i^e N_2(z) \Gamma_i h v_i \Delta v \right] \quad (5)$$

o índice *i* é o identificador da componente espectral dos canais utilizados para descrever a emissão espontânea e $\Delta v$ é o intervalo de frequências entre cada canal. As equações (4) e (5) foram resolvidas através do método APA que consiste em dividir a fibra em secções de comprimento reduzido, assumindo que a potência dos sinais não varia ao longo de cada secção. A cada secção é atribuído o valor médio da potência dos canais e, assim, é eliminada a dependência espacial das equações de taxa do modelo de Saleh:

$$P_k^{\pm out} = P_k^{\pm in} G_k(z) \quad (6)$$

$$P_i^{\pm out} = 2h v_i \Delta v \left( G_i(z-1) \right) \quad (7)$$

$$G_{k,i}(z) = \exp\left[ \left\{ \left(\sigma_{k,i}^e + \sigma_{k,i}^a\right) N_2(z) - \sigma_{k,i}^a \cdot (N_2 + N_1) \right\} z \cdot \Gamma_{k,i} - \alpha_{k,i} \right] \quad (8)$$

A variável $P_k^{\pm out}$ representa a potência ótica de saída de cada canal, cuja potência ótica de entrada é $P_k^{\pm in}$. $P_i^{\pm out}$ representa a potência ótica do ruído devido à emissão espontânea amplificada. $G_{ki}(z)$ representa o ganho em função da posição longitudinal na fibra, *z*. Considerando que a distribuição de potência ao longo da fibra é uniforme, aplicado o método APA para os sinais, resulta em [10]:

$$\langle P_k \rangle = P_k^{\pm in} \frac{(G_k - 1)}{\ln(G_k)} \quad (9)$$

O mesmo procedimento é realizado para a potência de ruído devido à emissão espontânea amplificada:

$$\langle P_i \rangle = 2h v_i \Delta v \cdot n_{sp} \left\{ \frac{(G_i - 1)}{\ln(G_i)} - 1 \right\} \quad (10)$$

sendo $n_{sp}$ o fator de inversão de população dado por:

$$n_{sp} = \frac{N_2}{N_2 \left(1 + \frac{\sigma_{k,i}^a}{\sigma_{k,i}^e}\right) - \frac{(N_2 + N_1)}{\sigma_{k,i}^e} \sigma_{k,i}^a - \frac{\alpha_{k,i}}{\Gamma_{k,i} \sigma_{k,i}^e}} \quad (11)$$

A densidade de população no estado excitado, $E_2$, é [8]:

$$N_2 = \frac{N_1 \left( \sum_{k=1}^{M} \frac{\langle P_k \rangle}{P_k^{sat}} + 2 \sum_{i=1}^{N} \frac{\langle P_i \rangle}{P_i^{sat}} \right)}{1 + \sum_{k=1}^{M} \frac{\langle P_k \rangle}{P_k^{sat}} \left(\frac{\sigma_{k,i}^e}{\sigma_{k,i}^a}\right) + 2 \sum_{i=1}^{N} \frac{\langle P_i \rangle}{P_i^{sat}} \left(\frac{\sigma_{k,i}^e}{\sigma_{k,i}^a} + 1\right)} \quad (12)$$

sendo a potência de saturação expressa da seguinte forma:

$$P_{ki}^{sat} = \frac{h\nu_{ki}\pi a^2}{\tau.\Gamma_{ki}\sigma_{ki}^a}. \quad (13)$$

Rearranjando (8), obtém-se o ganho de cada secção:

$$G_k = ex + \left\{ \left[ \frac{-(N_2+N_1)\Gamma_k\left(1+P_b-\frac{\sigma_{ki}^e}{\sigma_{ki}^a}P_a\right)}{1+P_a+P_b} - \alpha_k \right] \times L \right\} \quad (14)$$

onde $P_a$ e $P_b$ são variáveis intermédias, ajustadas iterativamente.

$$P_a = \sum_{k=1}^{M} \frac{P_k^{in}(G_k-1)}{P_k^{sat}\ln(G_k)} + 2\sum_{i=1}^{N} \frac{2h\nu_i\Delta\nu.n_{sp}}{P_k^{sat}} \left(\frac{(G_i-1)}{\ln(G_i)}-1\right) \quad (15)$$

$$P_b = \sum_{k=1}^{M} \frac{\sigma_{k,i}^e}{\sigma_{k,i}^a}\frac{P_k^{in}(G_k-1)}{P_k^{sat}\ln(G_k)} + 2\sum_{i=1}^{N} \frac{\sigma_{k,i}^e}{\sigma_{k,i}^a}\frac{2h\nu_i\Delta\nu.n_{sp}}{P_i^{sat}}\left(\frac{(G_i-1)}{\ln(G_i)}-1\right) \quad (16)$$

Esta aproximação permite um tempo de computação de aproximadamente 3 s para a simulação de um sistema com 10 canais num computador com 2 GB de RAM e um processador de 2.6 GHz dual core. Este valor para o tempo de computação é compatível com uma interface gráfica que permita atualizações em tempo real.

O modelo dinâmico implementado é uma versão melhorada do modelo de Sun proposta por Rieznik e Fragnito [11]. Nesse modelo, a dependência do ganho tem uma solução analítica se a população do estado excitado for constante ao longo da fibra, sendo necessário duas equações fundamentais, similares àquelas apresentadas anteriormente:

$$\frac{\partial N_2(z,t)}{\partial t} = -\frac{N_2(z,t)}{\tau} - \frac{1}{\rho A}\sum_{n=1}^{N}[(\gamma_n+\alpha_n)\times N_2(z,t)-\alpha_n]P_n(z,t) \quad (17)$$

$$\frac{\partial P_n(z,t)}{\partial z} = u_n[(\gamma_n+\alpha_n)N_2(z,t)-\alpha_n-\alpha_{loss}]\times P_n(z,t)+2\gamma_n\Delta\nu N_2(z,t) \quad (18)$$

onde $\gamma_n = \Gamma \times \sigma_k^e$ e $\alpha_n = \Gamma \times \sigma_k^a$. Como a maioria das situações simuladas corresponde a comprimentos de fibra ótica reduzidos e a sinais com potência reduzida (< 10 mW), nestas condições a atenuação podem ser consideradas desprezável. Assumindo que $N_2$ é constante ao longo da fibra e integrando em todo o seu comprimento, a solução da equação (18) é:

$$P_n^{out} = P_n^{in}(t)G_n(t) + 2n_{sp}\left[G_n(t)-1\right]\Delta\nu \quad (19)$$

onde

$$G_n(t) = \exp\{[(\gamma_n+\alpha_n)N_2(t)-\alpha_n]L\} \quad (20)$$

$$n_{sp} = \frac{N_2(t)\gamma_n}{(\gamma_n+\alpha_n)N_2(t)-\alpha_n} \quad (21)$$

sendo que (17) pode ser reescrita:

$$\frac{\partial N_2(z,t)}{\partial t} = -\frac{N_2(z,t)}{\tau} - \frac{1}{\rho SA}\sum_{n=1}^{N}\left[P_n^{out}(t)-P_n^{in}(t)-2\lambda_n\Delta\nu N_2\right](t)L \quad (22)$$

Esta última equação diferencial é resolvida usando a ferramenta do Matlab *ODE45*. As equações (19), (20) e (21) são determinadas para um dado período temporal, logo para a simulação de cenários de remoção/adição de canais numa rede ótica, é considerada a variação da potência ótica dos sinais roteados. O tempo de computação para este cenário de operação é de, aproximadamente, 1.9 s por equação num computador com 2 GB de RAM e um processador de 2.6 GHz *dual core*.

IV. Descrição das Interfaces Gráficas

A. Modelo estacionário

A interface visual para o modelo estacionário é composta por uma única janela com diferentes seções, figura 3. Nas seções do lado direito, estão todos os parâmetros editáveis e na secção do lado esquerdo, encontra-se a área de gráficos com os resultados da simulação.

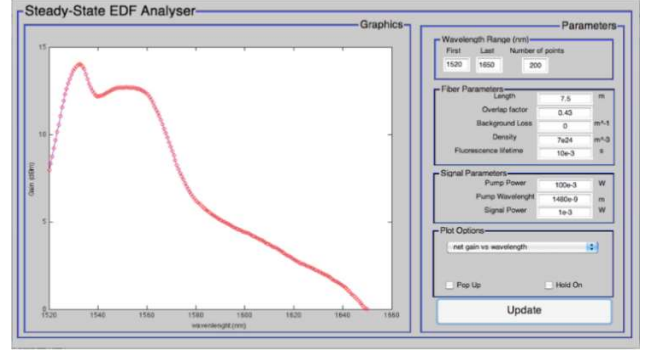

Figura 3. Imagem da interface gráfica do modelo estacionário.

Existem várias opções de funcionamento, sendo possível introduzir novas opções acedendo ao código fonte do programa, sendo um processo fácil e intuitivo que conta com os comentários e menus de ajuda do Matlab. Uma das opções, ativada na caixa denominada de "*hold on*", permite manter o gráfico atual e realizar novas simulações, tornando possível a comparação entre resultados. A opção "*pop up*" permite gerar os gráficos em novas janelas fora da interface.

A secção de parâmetros foi desenhada para ser de fácil utilização. Os parâmetros são todos editáveis e estão dispostos em três categorias. No canto superior direito encontra-se o menu de escolha da gama espectral de simulação e do número de pontos a simular. Por baixo dessa secção, está o espaço dedicado aos parâmetros físicos da fibra: comprimento, fator de sobreposição, atenuação perdas do meio, densidade de dopantes e tempo médio de vida dos dopantes no estado excitado. Por fim, encontramos uma secção dedicada aos parâmetros do sinal de bombeamento e dos sinais de dados a amplificar, sendo possível escolher o comprimento de onda e respetivo valor de potência, bem como a possibilidade do sinal de bombeamento ser co-propagante ou contra-propagante em relação ao sinal a amplificar. Em relação à área gráfica, as legendas dos gráficos são atualizadas automaticamente em função do tipo de gráfico escolhido, sendo feito o redimensionamento dos eixos quando o botão de "*update*" é pressionado.

O desempenho desta interface foi testada em vários cenários. Na figura 4 encontra-se o resultado referente ao ganho do AFDE em função da potência do sinal de bombeamento (1480 nm), para um único sinal de dados com um comprimento de onda de 1550 nm e uma potência ótica inicial de 1 mW. Foi considerado um valor para a densidade de portadores de $2\times10^{24}$ m$^{-3}$, um tempo médio de vida para o estado $^4I_{13/2}$ de 10 ms e um comprimento da fibra dopada de 10 m, sendo a atenuação considerada nula.

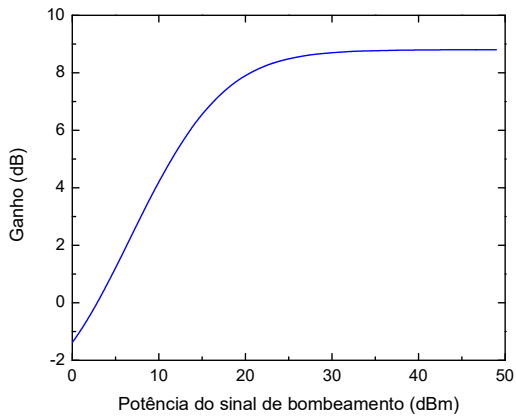

Figura 4. Ganho de um sinal com um comprimento de onda de 1550 nm em função da potência do sinal de bombeamento.

Observa-se que o ganho cresce linearmente com a potência do sinal de bombeamento até um valor de saturação. Com o aumento da potência de bombeamento, um maior número de portadores é excitado, produzindo emissão estimulada e, consequentemente, um valor de ganho superior. Quando todos os portadores ao longo da fibra estiverem excitados, o nível fundamental de energia está esgotado e um aumento da potência do sinal de bombeamento não produz qualquer efeito, mantendo-se o ganho constante. Este fenómeno de saturação do ganho é visível nos resultados da figura 4, para valores de potência do sinal de bombeamento superiores a 20 dBm.

Outra característica analisada por simulação foi a influência da potência do sinal de bombeamento na curva espectral de ganho, ilustrada na figura 5.

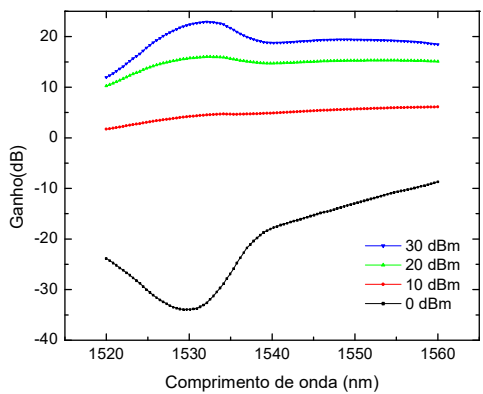

Figura 5. Espectro de ganho em função da potência do sinal de bombeamento.

Num AFDE, o comprimento da fibra é um parâmetro essencial. Em particular, para comprimentos elevados fibras (> 50 m), nos troços finais da fibra vai ocorrer absorção do sinal. Assim, um comprimento elevado da fibra dopada, aliado a um valor reduzido da potência do sinal de bombeamento, pode originar uma atenuação do sinal, enquanto que, quando a potência do sinal de bombeamento é elevada e a fibra dopada tem um comprimento reduzido, regista-se um uso ineficaz da energia disponível que não será absorvida ao longo de toda a extensão da fibra. Num exercício prático com a interface gráfica pode facilmente ilustrar−se o conceito de correlação de parâmetros. Como exemplo, foi simulado o impacte de dois parâmetros em simultâneo: o comprimento da fibra e a potência inicial dos sinais a amplificar, figura 6.

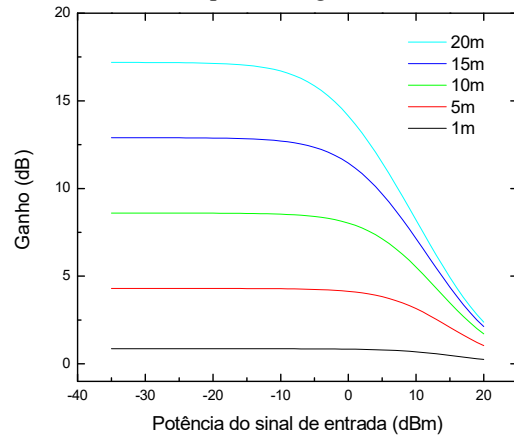

Figura 6. Ganho em função da potência do sinal de entrada para diversos valores do comprimento da fibra.

Para um valor da potência de bombeamento fixo, é disponibilizado um número fixo de portadores excitados e, consequentemente, existe uma limitação ao valor do ganho do amplificador. Para uma valor de potência do sinal de dados inferior a -10 dBm existem portadores excitados disponíveis para a realização do processo de emissão estimulada e, consequentemente de amplificação. Para potências superiores (> - 10 dBm), o número de portadores disponível decai rapidamente, atingindo-se uma situação de saturação da amplificação.

### B. Modelo dinâmico

A natureza dinâmica do modelo dinâmico do AFDE, requer que o algoritmo seja escrito por forma a resolver automaticamente o número necessário de equações, de acordo com as necessidades do utilizador. Este desafio é resolvido utilizando uma matriz que representa todas as operações de adição/remoção. Devido às limitações da ferramenta GUIDE do Matlab, foi estabelecido um compromisso para que a criação de uma interface pudesse demonstrar o conceito importante das operações de remoção e adição de canais, ou seja, o número máximo de canais e de operações foi limitado, respetivamente, a 8 e 4. A figura 7 mostra a imagem da interface para o modelo dinâmico.

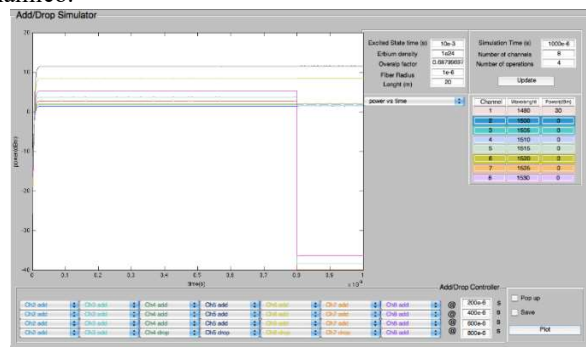

Figura 7. Imagem da interface gráfica do modelo dinâmico.

A interface possibilita o estudo de diversos cenários de remoção/remoção de canais numa rede com AFDE's. Tal como a interface usada no modelo estacionário, existe uma secção dedicada à edição de todos os parâmetros intrínsecos do AFDE. Numa secção próxima dos parâmetros editáveis, o utilizador define o período de tempo a simular, o número de canais e de operações a realizar, posteriormente, surgem as caixas necessárias para editar cada canal e operação. Existe uma secção onde os parâmetros de cada canal são introduzidos, bem como uma secção de controlo das operações de remoção/adição composta por menus de seleção, em que o utilizador escolhe se quer adicionar ou remover o canal respetivo de cada caixa.

O efeito da remoção de um canal num pequeno sistema de 3 canais é demonstrado na figura 8. A remoção da rede do canal com um comprimento de onda de 1555 nm origina um aumento da potência dos dois canais que permanecem na rede, com comprimentos de onda de 1545 nm e 1550 nm. O AFDE simulado tem 20 metros de fibra dopada com uma densidade de portadores de $10^{24}$ m$^{-3}$, distribuídos pela fibra segundo um fator de sobreposição de 0.68. A potência do sinal de bombeamento com um comprimento de onda de 1480 nm é de 100 mW.

A adição e remoção de canais alteram o ganho que o AFDE induz nos restantes canais e, consequentemente, ocorre a variação na potência dos restantes canais. Este efeito transiente tem origem na dinâmica de população do estado excitado. Quando um ou mais canais são removidos da rede isso significa um excedente de população de portadores excitados e um consequente aumento brusco da potência dos canais que permanecem na rede. O inverso acontece quando canais são adicionados causando um esgotamento brusco da população do estado excitado e, consequentemente, uma diminuição do ganho nos restantes canais.

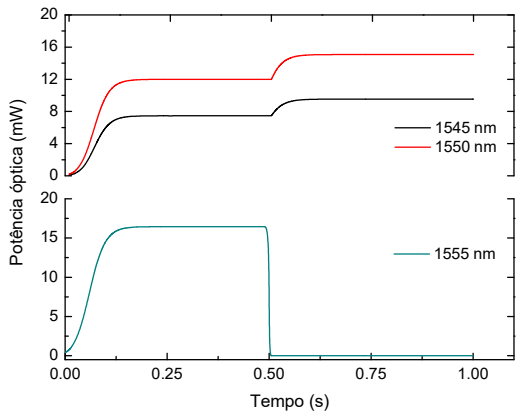

Figura 8. Potência ótica de saída do AFDE em função da potência dos canais amplificadores. Foram considerados 3 canais (1545, 1550 e 1555 nm), no instante 0.5 s é removido um canal da rede (1555 nm).

Numa situação real, a variação brusca da potência dos canais, causada pelo efeito transiente reduz a relação sinal/ruído, dificultando a sua deteção nos recetores óticos, representando um sério constrangimento ao uso de AFDEs nas redes de nova geração.

## V. CONCLUSÃO

Numa aproximação inovadora, foram desenvolvidas duas interfaces gráficas que permitem ilustrar os conceitos mais importantes associados ao desempenho de um AFDE. As interfaces, foram programadas em código aberto e podem ser modificadas consoante a necessidade dos seus utilizadores, apresentando tempos de computação reduzidos, facilidade de utilização e uma vasta gama de possibilidades a nível de simulação e estudo de diferentes cenários. A ferramenta proposta permite familiarizar utilizadores com as características e o comportamento do amplificador ótico de fibra dopada com iões Er$^{3+}$, sendo útil em ambiente de sala de aula ou em trabalhos realizados em períodos não letivos, sendo indubitavelmente uma mais-valia na relação entre ensino e aprendizagem.

O pacote de simulação pode ser descarregada através do endereço: http://www.lx.it.pt/~pandre/files/edfa.zip